\documentclass[preprint,showpacs,preprintnumbers,amsmath,amssymb]{revtex4} 
 
\usepackage{graphicx}
\usepackage{dcolumn}
\usepackage{bm}
 
\begin{document} 
 
\title{Relativistic mean-field approximation with 
density-dependent screening meson masses in nuclear matter} 
 
\author{Bao-Xi Sun$^{1,2}$, Xiao-Fu Lu$^{2,3,6}$, Peng-Nian Shen$^{6,1,2}$, 
En-Guang Zhao$^{2,4,5,6}$} 
 
\affiliation{${}^{1}$Institute of High Energy Physics, The Chinese Academy of 
Sciences, P.O.Box 918(4), Beijing  100039, China} 
 
\affiliation{${}^{2}$Institute of Theoretical Physics, The Chinese Academy of 
Sciences, Beijing  100080, China} 
 
\affiliation{${}^{3}$Department of Physics, Sichuan University, 
Chengdu  610064, China} 
 
\affiliation{${}^{4}$Center of Theoretical Nuclear Physics, 
National Laboratory of Heavy ion Accelerator,Lanzhou 730000,  China} 
 
\affiliation{${}^{5}$Department of Physics, Tsinghua University, 
Beijing 100084, China} 
 
\affiliation{${}^{6}$China Center of Advanced Science and Technology(World 
Laboratory), Beijing 100080, China} 
 
\begin{abstract} 
The Debye screening masses of the $\sigma$, $\omega$ and neutral 
$\rho$ mesons and the photon are calculated in the relativistic 
mean-field approximation. As the density of the nucleon increases, 
all the screening masses of mesons increase. It shows a different 
result with Brown-Rho scaling, which implies a reduction in the 
mass of all the mesons in the nuclear matter except the pion. 
Replacing the masses of the mesons with their corresponding 
screening masses in Walecka-1 model, five saturation properties of 
the nuclear matter are fixed reasonably, and then a 
density-dependent relativistic mean-field model is proposed 
without introducing the non-linear self-coupling terms of mesons. 
\end{abstract} 
\pacs{21.65.+f, 13.75.Cs}
\maketitle 

\newpage

\section{Introduction}

\par
In spite of the great success of the relativistic quantum field
theory(RQFT) in many areas of modern physics, some difficulties
have been found in applying this theory to nuclear systems. In
nuclear physics, the coupling constant of strong interaction
between nucleons is far larger than the fine structure constant in
quantum electrodynamics(QED). The ground state of the nuclear
matter or the finite nuclei is often defined as "vacuum", where
the Fermi sea is filled with interacting nucleons, and no
anti-nucleons and "holes" exist.

In 1970's, in order to solve the nuclear many-body problems with
RQFT, Walecka et al. developed the quantum hadrodynamics theory
(QHD)\cite{Wa.74,WS.86}. It is a important progress in nuclear
physics. Since then, the QHD theory has widely been used in
nuclear physics and nuclear astrophysics\cite{P.96,G.97}. However,
in the earliest QHD theory, the resultant compression modulus is
almost 550MeV\cite{Wa.74}, which is far from the experimental data
range of $200~-~300MeV$. To solve this problem, the nonlinear
self-coupling terms among $\sigma$ mesons were introduced in
addition to the mass term $\frac{1}{2} m^2_\sigma \sigma^2_{}$
\cite{BB.77},
\begin{equation}
U(\sigma)~=~\frac{1}{2} m^2_\sigma \sigma^2_{} ~+~\frac{1}{3} g_2
\sigma^3_{}~+~\frac{1}{4} g_3 \sigma^4_{}.
\end{equation}
Moreover, a self-coupling term of the vector meson
$$\frac{1}{4} c_3 (\omega^\mu\omega_\mu)^2$$
is added to produce the proper equation of state of nuclear
matter\cite{Bo.91,ST.94}. No doubt, additional parameters would
give more freedoms to fit the saturation curve of nuclear matter.
Zimanyi and Moszkowki developed the derivative scalar coupling
model yielding a compression modulus of 225MeV without any
additional parameter\cite{ZM.90}. Because all of these models
include the nonlinear density-dependent terms in the Lagrangian in
substance, the proper value of the compression modulus can be
obtained.

In the relativistic quantum field theory, the nonlinear
density-dependent terms in the Lagrangian represent higher-order
corrections. In this paper, we would calculate the self-energies
of virtual mesons with zero time-space momentum in the framework
of the relativistic mean-field approximation, and then the Debye
screening masses of the mesons would be discussed. Finally, we
would replace the masses of the mesons with their corresponding
screening masses in the Walecka-1 model, and develop
 a density-dependent relativistic mean-field theory without
additional parameters, which is similar to the density-dependent
quark-meson coupling model, i.e. quark-meson coupling
model-2(QMC-2)\cite{QMC.97}.

\section{The screening meson masses in the relativistic mean-field approximation}

the Lagrangian density in nuclear system can be written as
\begin{equation}
{\cal L}~=~{\cal L}_N~+~{\cal L}_{\sigma} ~+~{\cal L}_{\omega}
~+~{\cal L}_{\rho}~+~{\cal L}_{Int}.
\end{equation}
In this expression, the Lagrangian density for the free nucleon
field can be described by:
\begin{equation}
{\cal L}_N~=~\bar\psi \left(i\gamma_{\mu}\partial^{\mu}  -
M_N\right)\psi,
\end{equation}
where $\psi$ is the field of the nucleon and $M_N$ is the bare
mass of the nucleon. The free Lagrangian densities for the
$\sigma$, $\omega$, $\rho$ meson fields and the photon field can
be expressed by:
\begin{eqnarray}
{\cal L}_{\sigma}
&=&\frac{1}{2}\partial_\mu\sigma\partial^\mu\sigma-\frac{1}{2}
m^2_\sigma \sigma^2_{}, \\
 {\cal L}_{\omega}
&=&-\frac{1}{4}\omega_{\mu\nu}\omega^{\mu\nu}+ \frac{1}{2}
m^2_\omega\omega_\mu\omega^\mu , \\
 {\cal L}_{\rho}
&=&-{1\over4} \vec{R}_{\mu\nu} \cdot \vec{R}^{\mu\nu} +
{1\over2}m_\rho^{2} \vec{\rho}_\mu \cdot \vec{\rho}^\mu,  \\
 {\cal L}_{A}&=&-\frac{1}{4}F_{\mu\nu} F^{\mu\nu},
\end{eqnarray}
 where
\begin{eqnarray}
\omega_{\mu\nu}~&=&~\partial_\mu\omega_\nu-\partial_\nu\omega_\mu,
\\
\vec{R}_{\mu\nu}~&=&~\partial_\mu\vec{\rho}_\nu-\partial_\nu\vec{\rho}_\mu,
\\
F_{\mu\nu}~&=&~\partial_\mu A_\nu-\partial_\nu A_\mu
\end{eqnarray}
are corresponding field tensors, respectively. The interactive
Lagrangian density can be written as
\begin{eqnarray}
{\cal L}_{Int}&=& -g_\sigma\bar\psi\sigma\psi-g_\omega\bar\psi
\gamma_\mu \omega^\mu \psi \nonumber \\
&&-~g_{\rho}  \bar\psi \gamma^\mu \frac{\vec{\tau}}{2} \cdot
\vec{\rho}_\mu \psi -e\bar\psi \gamma_\mu \frac{1+\tau_3}{2} A^\mu
\psi
\end{eqnarray}
with $\vec{\tau}$ being the Pauli matrix.

In the framework of relativistic mean-field approximation, the
meson fields operators and the electromagnetic field operator can
be replaced by their expectation values in the nuclear matter
\cite{WS.86}:
\begin{eqnarray}
\sigma&\rightarrow&\langle\sigma\rangle~=~\sigma_0, \\
\omega_\mu&\rightarrow&\langle\omega_\mu\rangle~=~\omega_0\delta^{0}_{\mu},\\
\vec{\rho}_\mu&\rightarrow&\langle\vec{\rho}_\mu\rangle~=~\rho_0\delta^{0}_{\mu},\\
A_\mu&\rightarrow&\langle A_\mu \rangle~=~A_0\delta^{0}_{\mu},
\end{eqnarray}
where $\sigma_0$ is the expectation value of the scalar meson
field operator, $\omega_0$ and $\rho_0$ are the expectation values
of the time-like components of $\omega$ meson and neutral $\rho$
meson field operators, respectively, and $A_0$ denotes the scalar
potential of the electromagnetic field in the nuclear system. The
Lagrangian density for nuclear matter in the relativistic
mean-field approximation then reads
\begin{eqnarray}
{\cal L}_{RMF}&=& \bar\psi \left(i\gamma_{\mu}\partial^{\mu}  -
M_N \right)\psi -g_\sigma\bar\psi\psi\sigma_{0}
-g_\omega\bar\psi\gamma^{0}\psi\omega_{0}
-g_{\rho} \bar\psi \gamma^0 \frac{\tau_3}{2} \psi  \rho_{0} \nonumber\\
&&-\frac{1}{2} m^2_\sigma \sigma^2_{0}  +\frac{1}{2}
m^2_\omega\omega_{0}^2 + {1\over2}m_\rho^{2} \rho_{0}^2.
\end{eqnarray}
In the nuclear matter, because of the long-range Coulomb repulsive
interaction between protons, the nuclear matter would not be
bound. As a model of nuclear matter, the electromagnetic
interaction usually be ignored. However, in the finite nuclei, the
electromagnetic interaction should be considered in the structure
calculation.

The screening mass of the meson $m^{\ast}_\alpha$ is defined from
the static infrared limit $k^0=0$ and then $\vec{k}\rightarrow 0$
as Ref. \cite{SRK.95}
\begin{equation}
\label{eq:Debye-M}
m^\ast_\alpha~=~\sqrt{m^2_\alpha~+~\Sigma_\alpha(k=0, \rho_p,
\rho_n)},~~~~~~(\alpha~=~\sigma, \omega, \rho_0).
\end{equation}
It represents the inverse Debye screening length and implies the
long-distance correlations. In the following, we will calculate
the self-energy of virtual mesons with $k=0$ in the framework of
relativistic mean-field approximation, and then the screening
masses of the mesons and the photon are obtained.

The interactive perturbation Hamiltonian of the model can be
expressed as
\begin{equation}
{\cal H}_I~=~ g_\sigma\bar\psi\sigma\psi+g_\omega\bar\psi
\gamma_\mu \omega^\mu \psi 
+~g_{\rho}  \bar\psi \gamma^\mu \frac{\vec{\tau}}{2} \cdot
\vec{\rho}_\mu \psi +e\bar\psi \gamma_\mu \frac{1+\tau_3}{2} A^\mu
\psi,
\end{equation}
and the S-matrix can be written as
\begin{equation}
\hat{S} ~=~\hat{S}_0~+~\hat{S}_1~+~\hat{S}_2~+~\ldots,
\end{equation}
where
\begin{equation}
\hat{S}_n ~=~\frac{(-i)^n}{n !}\int d^{4}x_1 \int d^{4}x_2 \ldots
\int d^{4}x_n   T \left[{\cal H}_I(x_1){\cal H}_I(x_2) \ldots
{\cal H}_I(x_n) \right].
\end{equation}
In the 2nd-order approximation, only
\begin{equation}
\label{eq:S2} \hat{S}_2 ~=~\frac{(-i)^2}{2 !}\int d^{4}x_1 \int
d^{4}x_2
 T\left[{\cal H}_I(x_1){\cal H}_I(x_2) \right].
\end{equation}
should be calculated.

The nucleon field operator $\psi(x)$ and its conjugate operator
$\bar\psi(x)$ can be expanded in terms of a complete set of
solutions of the Dirac equation:
\begin{eqnarray}
\label{eq:pfield}
 \psi(\vec{x},t) &=
&\sum_{\eta=1,2}
\sum_{\lambda=1,2}\int\frac{d^{3}p}{(2\pi)^{\frac{3}{2}}}\sqrt{\frac{M^{\ast}_N}{E^{\ast}(p)}}
 \nonumber \\
&&\left[ A_{p\lambda}U_\eta(p,\lambda)\exp\left(i\vec{p}\cdot\vec{x}
-i\varepsilon^{(+)}(p)t\right)\right. \nonumber \\
&+&\left.
B^{\dagger}_{p\lambda}V_\eta(p,\lambda)\exp\left(-i\vec{p}\cdot\vec{x}-
i\varepsilon^{(-)}(p)t\right)\right],
\end{eqnarray}
\begin{eqnarray}
\label{eq:bpfield}
 \bar\psi(\vec{x},t) &=&\sum_{\eta=1,2}
 \sum_{\lambda=1,2}
\int\frac{d^{3}p}{(2\pi)^{3/2}}\sqrt{\frac{M^{\ast}_N}{E^{\ast}(p)}} \nonumber \\
&&\left[A_{p\lambda}^{\dagger}\bar{U}_\eta(p,\lambda)
\exp\left(-i\vec{p}\cdot\vec{x}+i\varepsilon^{(+)}(p)t\right)\right. \nonumber \\
&+&\left.B_{p\lambda}\bar{V}_\eta(p,\lambda)
\exp\left(i\vec{p}\cdot\vec{x}+i\varepsilon^{(-)}(p)t\right)\right],
\end{eqnarray}
where $U_\eta(p,\lambda)$ and $V_\eta(p,\lambda)$ are Dirac
spinors for the positive and negative energies, respectively, and
\begin{equation}
\label{eq:equal}
 \sum_{\lambda=1,2}U_\eta(p,\lambda)
\bar{U}_{\eta^\prime}(p,\lambda)~=~\left(\frac{\rlap{/} p +
M^{\ast}_N}{2M^{\ast}_N}\right) \delta_{\eta \eta^\prime},
\end{equation}
\begin{equation}
 \sum_{\lambda=1,2}V_\eta(p,\lambda)
\bar{V}_{\eta^\prime}(p,\lambda)~=~\left(\frac{\rlap{/} p -
M^{\ast}_N}{2M^{\ast}_N}\right) \delta_{\eta \eta^\prime},
\end{equation}
with $\lambda$ and $\eta$ denoting the spin and isospin for the
nucleon, and $M^{\ast}_N~=~M_N~+~g_{\sigma} \sigma_0 $ is the
effective mass of the nucleon.
\begin{equation}
\varepsilon^{(\pm)}(p)~=~{\pm} E^{\ast}(p)~+~g_\omega
\omega_0~+~g_\rho \frac{\tau_3}{2}\rho_0~+~e \frac{1+\tau_3}{2}
A_0
\end{equation}
with $E^{\ast}(p)=\sqrt{\vec{p}^2~+~{M_{N}^\ast}^2}$, are the
positive and negative energy eigenvalues for the Dirac equation of
the nucleon in the nuclear matter, respectively.
Assuming there are no antinucleons in the nuclear matter or finite
nuclei, only positive-energy components are considered in
Eqs.~(\ref{eq:pfield}) and ~(\ref{eq:bpfield}).

When a scalar meson of momentum $k$ is considered in nuclear
matter, the scalar meson field operator $\sigma(k,x)$ can be
expressed as
\begin{equation}
\sigma(k,x)~=~ a(k)\exp(-ik \cdot x) +a^\dagger(k)\exp(ik \cdot
x).
\end{equation}
If the coupling constants $g_\sigma$, $g_\omega$ and $g_\rho$, the
 proton charge $e$, and the masses of mesons are supposed have
 already been renormalized, the contribution of a single nucleon loop
is not necessarily calculated\cite{Bj.64}, only these
contributions from the Feynman diagrams in Fig.1 should be
considered.

The expectation value of $\hat{S}_2$ can be written as
\begin{eqnarray}
\label{eq:S2-1} \langle~k_2~|~\hat{S}_2~|~ k_1~\rangle
~&=&~-ig_\sigma^2(2\pi)^4 \delta^4(p_1+k_1-p_2-k_2)
 \nonumber\\
&&\sum_{\eta=1,2}\sum_{\lambda=1,2}\int\frac{d^3p}{(2\pi)^3}
\frac{M^{\ast}_N}{E^{\ast}(p)}\theta(p_F-|\vec{p}|) \\
&&\bar{U}_\eta(p,\lambda) \left(\frac{1}{\rlap{/} p -\rlap{/} k -
M^{\ast}_N}~+~ \frac{1}{\rlap{/} p +\rlap{/} k -
M^{\ast}_N}\right)
 U_\eta(p,\lambda), \nonumber
\end{eqnarray}
where $k_1=k_2=k, $ and $p_1=p_2=p=(E^{\ast}(p),\vec{p}), $ and
$\theta(x)$ is the step function.

Considering the diagrams in Fig.1, the scalar meson propagator
$G(k)$ in nuclear matter can be derived as
\begin{eqnarray}
G(k)~&=&~\frac{1}{(2\pi)^4}\frac{i}{k^2-m^2_\sigma+i\varepsilon}+
\frac{1}{(2\pi)^4}\frac{i}{k^2-m^2_\sigma+i\varepsilon}   \nonumber \\
&&\sum_{\eta=1,2} \sum_{\lambda=1,2} (-ig_\sigma^2)(2\pi)^4
\int\frac{d^3p}{(2\pi)^3}
\frac{M^\ast_N}{E^\ast(p)}\theta(p_F-|\vec{p}|) \nonumber \\
&&\bar{U}_\eta(p,\lambda) \left(\frac{1}{\rlap{/} p -\rlap{/} k -
M^{\ast}_N}~+~ \frac{1}{\rlap{/} p +\rlap{/} k -
M^{\ast}_N}\right)
 U_\eta(p,\lambda)
\frac{1}{(2\pi)^4}\frac{i}{k^2-m^2_\sigma+i\varepsilon} \nonumber \\
~&=&~\frac{i}{(2\pi)^4}
\frac{1}{k^2-m^2_\sigma+i\varepsilon}+\frac{i}{(2\pi)^4}
\frac{g_\sigma^2}{k^2-m^2_\sigma+i\varepsilon}\sum_{\eta=1,2}
\sum_{\lambda=1,2} \int\frac{d^3p}{(2\pi)^3}
\frac{M^{\ast}_N}{E^{\ast}(p)}\theta(p_F-|\vec{p}|) \nonumber \\
&&\bar{U}_\eta(p,\lambda) \left(\frac{1}{\rlap{/} p -\rlap{/} k -
M^{\ast}_N}~+~ \frac{1}{\rlap{/} p +\rlap{/} k -
M^{\ast}_N}\right)
 U_\eta(p,\lambda)
\frac{1}{k^2-m^2_\sigma+i\varepsilon}.
\end{eqnarray}
According to the Dyson equation
\begin{equation}
\frac{i}{k^2-m^{2}_{\sigma}-\Sigma_\sigma+i\varepsilon}
~=~\frac{i}{k^2-m^2_\sigma+i\varepsilon}
+\frac{i}{k^2-m^2_\sigma+i\varepsilon} ~ \Sigma_\sigma ~
\frac{1}{k^2-m^2_\sigma+i\varepsilon},
\end{equation}
we obtain the self-energy of the scalar meson in the nuclear
matter
\begin{eqnarray}
\Sigma_\sigma~&=&~ g_\sigma^2~\sum_{\eta=1,2}\sum_{\lambda=1,2}
\int\frac{d^3p}{(2\pi)^3}
\frac{M^{\ast}_N}{E^{\ast}(p)}\theta(p_F-|\vec{p}|) \nonumber \\
&&\bar{U}_\eta(p,\lambda) \left(\frac{1}{\rlap{/} p -\rlap{/} k -
M^{\ast}_N}~+~ \frac{1}{\rlap{/} p +\rlap{/} k -
M^{\ast}_N}\right)
 U_\eta(p,\lambda),
\end{eqnarray}
where
\begin{eqnarray}
\label{eq:deduce} &&\sum_{\lambda=1,2}\bar{U}_\eta(p,\lambda)
\left(\frac{1}{\rlap{/} p -\rlap{/} k - M^{\ast}_N}~+~
\frac{1}{\rlap{/} p +\rlap{/} k - M^{\ast}_N}\right)
 U_\eta(p,\lambda) \nonumber \\
&=&\sum_{\lambda=1,2}Tr\left[ \left(\frac{1}{\rlap{/} p -\rlap{/}
k - M^{\ast}_N}~+~ \frac{1}{\rlap{/} p +\rlap{/} k -
M^{\ast}_N}\right)
 U_\eta(p,\lambda)\bar{U}_\eta(p,\lambda)\right] \nonumber \\
&=&Tr\left[ \left(\frac{\rlap{/} p -\rlap{/} k + M^{\ast}_N}{ (p -
k)^2 - {M^{\ast}_N}^2}~+~ \frac{\rlap{/} p +\rlap{/} k +
M^{\ast}_N}{ (p + k)^2 - {M^{\ast}_N}^2}\right) \frac{\rlap{/} p +
M^{\ast}_N}{2M^{\ast}_N}\right].
\end{eqnarray}
In the above deduction, Eq.~(\ref{eq:equal}) is used.

In order to obtain the Debye screening mass of the scalar meson,
we calculate the self-energy of the scalar meson in the nuclear
matter in the static infrared limit $k^0=0$ and then
$\vec{k}\rightarrow0$. Because the time-space momentum of virtual
scalar meson is zero, $k=0$, we have $k^2=0$. Considering
$p^2={M^{\ast}_N}^2$, the virtual scalar meson self-energy
\begin{eqnarray}
\Sigma_\sigma~&=&~ g_\sigma^2~\sum_{\eta=1,2}
\int\frac{d^3p}{(2\pi)^3}
\frac{M^{\ast}_N}{E^{\ast}(p)}\theta(p_F-|\vec{p}|) \nonumber \\
&& Tr\left[ \left(\frac{\rlap{/} p -\rlap{/} k + M^{\ast}_N}{ -2p
\cdot k }~+~ \frac{\rlap{/} p +\rlap{/} k + M^{\ast}_N}{ 2p \cdot
k}\right) \frac{\rlap{/} p +
M^{\ast}_N}{2M^{\ast}_N}\right] \nonumber \\
~&=&~ g_\sigma^2~\sum_{\eta=1,2} \int\frac{d^3p}{(2\pi)^3}
\frac{M^{\ast}_N}{E^{\ast}(p)}\theta(p_F-|\vec{p}|)
 Tr\left[ \frac{\rlap{/} k}{p \cdot
k} \cdot \frac{\rlap{/} p + M^{\ast}_N}{2M^{\ast}_N}\right] \nonumber \\
~&=&~\frac{g_\sigma^2}{M^{\ast}_N}\sum_{\eta=1,2}\frac{2}{(2\pi)^3}
\int d^3p
\frac{M^{\ast}_N}{E^{\ast}(p)}\theta(p_F-|\vec{p}|) \nonumber\\
~&=&~\frac{g_\sigma^2 (\rho_S^p~+~\rho_S^n)}{M^{\ast}_N},
\end{eqnarray}
where $\rho_S^p$ and $\rho_S^n$ are the scalar densities of
protons and neutrons with
\begin{eqnarray}
\rho_S^p&=&
 2
 \int_{p}\frac{d^3p}{(2\pi)^3}\frac{M^{\ast}_N}{(\vec{p}^2+{M^{\ast}_N}^2)^\frac{1}{2}},\\
\rho_S^n&=&
 2 \int_{n}\frac{d^3p}{(2\pi)^3}\frac{M^{\ast}_N}{(\vec{p}^2+{M^{\ast}_N}^2)^\frac{1}{2}}.
\end{eqnarray}

Similarly, the self-energies  of the $\omega$, neutral $\rho$ and
$\gamma$ with $k=0$ may be obtained:
\begin{eqnarray}
\Sigma_\omega&=&\frac{g_\omega^2
(\rho_S^p~+~\rho_S^n)}{2M^{\ast}_N}, \\
\Sigma_\rho&=&\frac{g_\rho^2
(\rho_S^p~+~\rho_S^n)}{8M^{\ast}_N}, \\
\Sigma_A&=&\frac{e^2 \rho_S^p}{2M^{\ast}_N}.
\end{eqnarray}

According to Eq.~(\ref{eq:Debye-M}), the Debye screening masses of
the $\sigma$, $\omega$ neutral $\rho$ and the photon can be
written as
\begin{equation}
\label{eq:msig}
m^\ast_\sigma(\rho_N,f_p)=\sqrt{m^2_\sigma~+~\Sigma_\sigma},
\end{equation}
\begin{eqnarray}
m^\ast_\omega(\rho_N,f_p)&=&\sqrt{m^2_\omega~+~\Sigma_\omega},  \\
m^\ast_\rho(\rho_N,f_p)&=&\sqrt{m^2_\rho~+~\Sigma_\rho}, \\
m^\ast_A(\rho_N,f_p)&=&\sqrt{\Sigma_A}=\sqrt{\frac{e^2
\rho_S^p}{2M^{\ast}_N}}.
\end{eqnarray}
They are also the functions of the density of nuclear matter
$\rho_N$ and the proton fraction $f_p=Z/A$.

In the relativistic mean-field approximation,
\begin{equation}
m^2_\sigma \sigma_0~=~-g_\sigma (\rho_S^p~+~\rho_S^n),
\end{equation}
so the effective mass of the nucleon can be written as
\begin{eqnarray}
M^\ast_N&=&M_N~+~g_\sigma \sigma_0 \nonumber \\
&=&M_N~-~\frac{g^2_\sigma(\rho_S^p~+~\rho_S^n)}{m^2_\sigma}.
\end{eqnarray}
From Eq.~(\ref{eq:msig}), the screening mass of the scalar meson
can be expanded as
\begin{equation}
m^\ast_\sigma~=~m_\sigma\left(1+\frac{1}{2} \cdot
\frac{g^2_\sigma}{m^2_\sigma} \cdot
\frac{\rho_S^p~+~\rho_S^n}{M^{\ast}_N}+... \right).
\end{equation}
To the first order of the scalar density of the nucleon,
\begin{equation}
\frac{\delta m_\sigma}{m_\sigma}/\frac{\delta
M_N}{M_N}~=~-\frac{1}{2} \cdot \frac{M_N}{M^{\ast}_N},
\end{equation}
where $\delta m_\sigma=m^{\ast}_\sigma-m_\sigma$, and $\delta
M_N=M^{\ast}_N-M_N$, At the low density of the nuclear matter,
$M^{\ast}_N \approx M_N$, and
\begin{equation}
\frac{\delta m_\sigma}{m_\sigma}/\frac{\delta
M_N}{M_N}~\approx~-\frac{1}{2}.
\end{equation}
Obviously, the fractional change of the mass of the scalar meson
is -1/2 times that of the nucleon in the nuclear matter, and this
simple value of the ratio is similar to the scaling relation in
Ref.~\cite{QMC.97}.

The ratio of the effective mass to the bare mass for the nucleon
and the ratios of screening mass to the bare mass for mesons as
the functions of the nucleon density for the symmetric nuclear
matter are shown in Fig.2. It is seen that when the number density
of nucleon increases, the effective mass of the nucleon decreases,
and the screening masses of mesons increase. The increase of
screening masses of the mesons means that as the density of the
nuclear matter become dense, the range of nuclear forces
decreases, and the Debye screening effect enhances in the denser
nuclear matter. Our results on the screening masses of the mesons
are different from the Brown-Rho scaling\cite{Br.91}, in which the
 masses of the mesons are reduced in the nuclear matter,
similarly to the reduction of the effective mass of the nucleon,
as the density of the nuclear matter increases.

\section{The equation of state for nuclear matter}

The relativistic mean-field results may be derived by summing the
tadpole diagrams self-consistently in nuclear matter, retaining
only the contributions from nucleons in the filled Fermi sea in
the evaluation of the self-energy and energy density\cite{WS.86},
so the relativistic mean-field approximation is consistent to the
relativistic Hartree approximation in the calculation of nuclear
matter. Because the time-space momentum $k$ of virtual mesons in
the tadpole diagrams is zero approximately, if the Debye screening
effect is considered, the masses of mesons in the relativistic
mean-field approximation should be replaced by their screening
masses, respectively. Therefore, the total energy density and the
pressure of nuclear matter can be deduced to
\begin{equation}
\varepsilon = \frac{1}{2} {m^{\ast}}^2_\sigma \sigma^2_{0} -
\frac{1}{2} {m^{\ast}}^2_\omega \omega^2_{0} - \frac{1}{2}
{m^{\ast}}^2_\rho \rho^2_{0}
 + \sum_{B=p,n} \varepsilon_B,
\end{equation}
and
\begin{eqnarray}
p~&=&~\frac{-1}{3}\sum_{B=p,n} \left(
 -\varepsilon_B + M_{N}^{\ast} \rho_{S}^B
+ g_{\omega}  \omega_{0} \rho_{V}^B +  g_{\rho} \frac{\tau_3}{2}
\rho_{0} \rho_{V}^B
 \right) \nonumber \\
&& - \frac{1}{2} {m^{\ast}}^2_\sigma \sigma^2_{0}
 + \frac{1}{2} {m^{\ast}}^2_\omega \omega^2_{0} + \frac{1}{2} {m^{\ast}}^2_\rho
 \rho^2_{0},
\end{eqnarray}
with
\begin{equation}
\varepsilon_B = \frac{2}{\left( 2\pi \right)^3}
\int^{p_{F}(B)}_{0} d\vec{p} \left( \left(\vec{p}^2 +
{M_{N}^{\ast}}^2 \right)^{\frac{1}{2}} + \left(  g_{\omega}
\omega_0 + g_{\rho} \frac{\tau_3}{2} \rho_{0} \right) \right),
\end{equation}
and the vector densities of protons and neutrons being
\begin{eqnarray}
\rho_V^p&=&
 2
 \int_{p}\frac{d^3p}{(2\pi)^3},\\
\rho_V^n&=&
 2 \int_{n}\frac{d^3p}{(2\pi)^3},
\end{eqnarray}
respectively.

By fitting the saturation properties of nuclear matter, the
parameters of the relativistic mean-field approximation in which
the mesons have density-dependent screening masses can be fixed
\begin{equation}
\frac{g^2_\sigma}{m^2_\sigma}~=~8.297fm^2,~~~~~~
\frac{g^2_\omega}{m^2_\omega}~=~3.683fm^2,~~~~~~
\frac{g^2_\rho}{m^2_\rho}~=~5.187fm^2.
\end{equation}
With these parameters  we obtain a saturation density of
$0.149\mbox{fm}^{-3}$, a binding energy of $16.669$~MeV, a
compression modulus of $280.1$~MeV, a symmetry energy coefficient
of $32.8$~MeV and an effective nucleon mass of $0.808M_N$ for the
symmetric nuclear matter.

The saturation curves for the nuclear matter with different
parameter sets are plotted in Fig.3. Comparing these results with
each other, we see that when the density of the nucleon becomes
larger than the saturation density of nuclear matter, the average
energy per nucleon with our model parameter set increases more
slowly than those with NL3\cite{P.97}, NLSH\cite{P.93} and
TM1\cite{ST.94} . It manifests that the equation of state for
nuclear matter in our model is softer than those in the other
models, because the mesons have density-dependent screening
masses.

Due to the approximate chiral symmetry restoration in nuclear
matter, the effective nucleon mass is reduced. Brown-Rho scaling
would imply a similar reduction in the mass of all the mesons
except the pion. However, this scaling would not lead to
reasonable equation of state for the nuclear matter. Our model
gives different results with Brown-Rho scaling, and can fit the
correct saturation properties of nuclear matter in the framework
of relativistic mean-field approximation.

In the nuclear matter, The screening mass of photon is only
related to the scalar density of the proton, but not the momentum
of the photon. In a symmetric nuclear matter, if one takes a
nucleon number density of $0.16fm^{-3}$, and the bare nucleon
mass, the resultant screening mass of photon is about $5.42MeV$
\cite{Sun.02}. Although the photon gains an screening mass in the
nuclear matter, the range of Coulomb repulsive force is still
large enough, and the nuclear matter would not be bound as the
electromagnetic interaction is considered. In the case of finite
nuclei, the contribution from the photon mass term should be
included in the relativistic Hartree approximation or relativistic
mean-field approximation.

In nuclear matter, the screening masses of the mesons increase
with the density of the nuclear matter. This result is equivalent
to the statement that the coupling constants decrease with the
increasing number density of nucleon when the masses of mesons
retain constant. At this point, our model is consistent to the
model in Ref.~\cite{Toki.92}.

\section{Summary}

\par
In summary, we calculate the self-energies of virtual mesons with
zero time-space momentum in the relativistic mean-field
approximation, and the Debye screening masses of the mesons are
obtained, which  are the functions of the scalar densities of
protons and neutrons. The screening masses of the mesons increase
with the density of the nucleon. It shows different results with
Brown-Rho scaling. Replacing the masses of mesons with
corresponding screening masses in the relativistic mean-field
approximation, we obtain a relativistic density-dependent nuclear
model. In this model, the nonlinear self-coupling terms of mesons
are not needed and only three model parameters are required. With
this model, five saturation properties of symmetric nuclear
matter, the saturation density, the binding energy, the effective
nucleon mass, the compression modulus and the asymmetric energy,
are calculated. In denser nuclear matter, the equation of state
with this model is softer than those of previous models. This
implies that the dense matter in the core of neutron stars or the
center of the relativistic heavy ion collision might be described
correctly.

This work was supported in part by the Major State Basic Research
Development Programme under Contract No. G2000-0774, the CAS
Knowledge Innovation Project No. KJCX2-N11 and the National
Natural Science Foundation of China under grant numbers 10075057
and 90103020.

\newpage

\begin{figure*}
\includegraphics{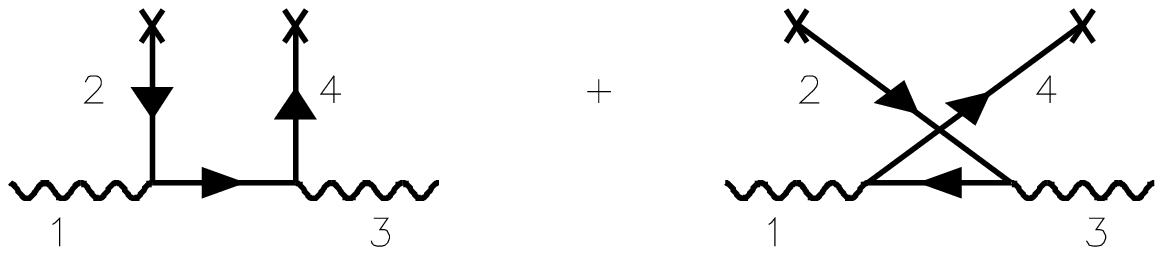}
\caption{\label{fig1}Feynman diagrams for the meson or the photon self-energy in 
nuclear matter, while 1 and 2 denote particles of the initial 
state, 3 and 4 denote particles of the final state.}
\end{figure*}

\begin{figure*}
\includegraphics{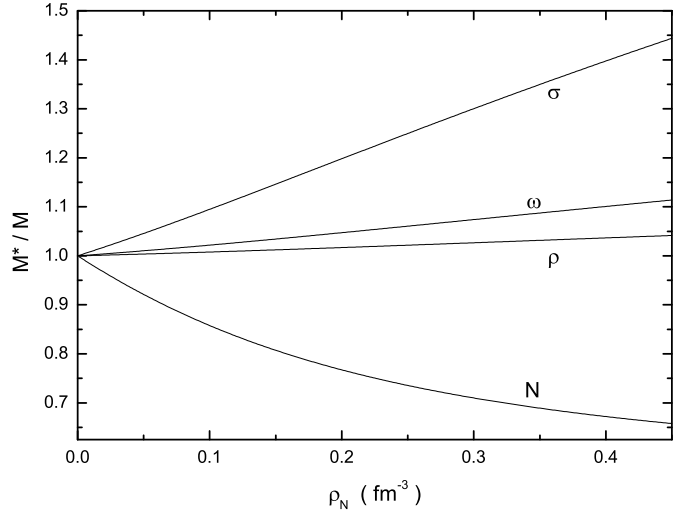}
\caption{\label{fig2}The ratios of the effective or screening mass to the bare mass 
 $M^\ast/M$ for the nucleon and the 
mesons as a function of the nucleon density $\rho_N$ in the 
symmetric 
nuclear matter in this model.}
\end{figure*}

\begin{figure*}
\includegraphics{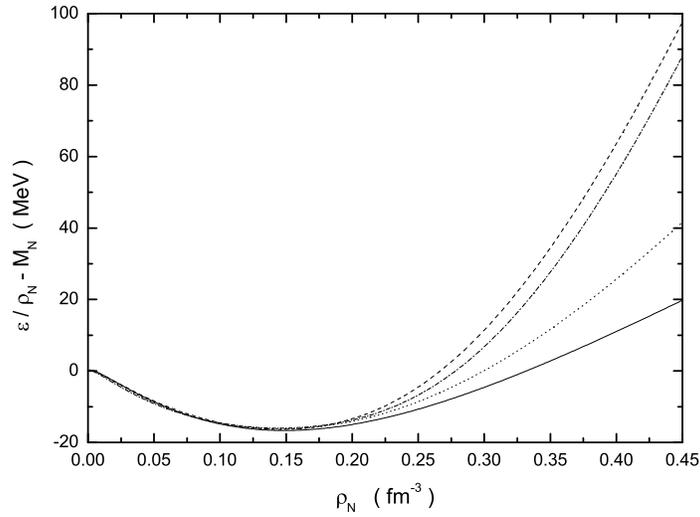}
\caption{\label{fig3}Average energy per nucleon $\varepsilon/\rho_N~-~M_N$ 
as a function of nucleon density $\rho_N$ with the different 
parameter sets. The solid line denotes the curve obtained in this 
model, the dash-dot line is for the NL3 parameters\cite{P.97}, the 
dash line is for the NLSH parameters\cite{P.93}, and the dot line 
is for the TM1 parameters\cite{ST.94}.}
\end{figure*}

\end{document}